\def\beq#1{\begin{equation}\label{#1}}
\def\eeq{\end{equation}}
\def\beqa#1{\begin{eqnarray}\label{#1}}
\def\eeqa{\end{eqnarray}}

\documentclass[12pt,preprint]{aastex}

\usepackage{emulateapj5}

\usepackage{graphicx}
\usepackage{epsfig}

\newcommand{\be}{\begin{equation}}
\newcommand{\ee}{\end{equation}}
\newcommand{\ba}{\begin{eqnarray}}
\newcommand{\ea}{\end{eqnarray}}

\shorttitle{Photometric Redshift Estimator for SNe Ia}
\shortauthors{Wang}

\begin{document}

\title{A Model-Independent Photometric Redshift Estimator 
for Type Ia Supernovae}
\author{Yun~Wang$^{1}$}
\altaffiltext{1}{Homer L. Dodge Department of Physics \& Astronomy, 
			Univ. of Oklahoma,
                 440 W Brooks St., Norman, OK 73019;
                 email: wang@nhn.ou.edu}

\begin{abstract}

The use of type Ia supernovae (SNe Ia) as cosmological standard candles
is fundamental in modern observational cosmology.
In this letter, we derive a simple empirical photometric redshift estimator 
for SNe Ia using a training set of SNe Ia with multiband ($griz$) light-curves
and spectroscopic redshifts obtained by the Supernova Legacy Survey (SNLS).
This estimator is analytical and model-independent; it does not use spectral 
templates. We use all the available SNe Ia from SNLS with near maximum
photometry in $griz$ (a total of 40 SNe Ia) to train and test our 
photometric redshift estimator.
The difference between the estimated redshifts $z_{phot}$ and the 
spectroscopic redshifts $z_{spec}$, $(z_{phot}-z_{spec})/(1+z_{spec})$, 
has rms dispersions of 0.031 for 20 SNe Ia used in the training set, 
and 0.050 for 20 SNe Ia not used in the training set. 
The dispersion is of the same order of magnitude as the flux uncertainties 
at peak brightness for the SNe Ia. There are no outlyers.

This photometric redshift estimator should significantly enhance the
ability of observers to accurately target high redshift SNe Ia for spectroscopy
in ongoing surveys. It will also dramatically boost the cosmological impact 
of very large future supernova surveys, such as those planned for 
Advanced Liquid-mirror Probe for Astrophysics, Cosmology and Asteroids (ALPACA),
and the Large Synoptic Survey Telescope (LSST).

\end{abstract}



\section{Introduction}

The use of type Ia supernovae (SNe Ia) as cosmological standard candles
\citep{Phillips93,Riess95,Wang03}
is fundamental in modern observational cosmology. Obtaining the spectroscopic
redshifts of SNe Ia is the most costly aspect of supernova surveys.

The use of broadband photometry in multiple filters to estimate redshifts
of galaxies has become well established \citep{Weymann99}. 
There are two different approaches
in estimating photometric redshifts of galaxies.
In the empirical fitting method \citep{Connolly95,Wang98},
a training set of galaxies with measured spectroscopic redshifts
are used to derive analytical formulae relating the redshift to
colors and magnitudes.
In the template fitting technique (see for example,
\cite{Puschell82,Lanzetta96,Mobasher96,Sawicki97}),
the observed colors are compared with the predictions of a set of
galaxy SED templates.   

The multiband photometry of SNe has been used to select SNe Ia
candidates \citep{Dahlen02,Riess04,Johnson06}. \cite{Strolger04}
used the template fitting method to estimate the photometric redshift
of SN host galaxies.

\cite{Cohen00} carried out a blind test of the predictions of 
\cite{Wang98} for galaxies in the Hubble Deep Field North (HDF), 
and demonstrated that this technique is capable of reaching
a precision of $\sigma[(z_{phot}-z_{spec})/(1+z_{spec})]=0.05$ for the 
majority of galaxies with $z<1.3$.
In this paper, we modify and further develop the technique
of \cite{Wang98}, to derive a simple and model-independent
empirical photometric redshift estimator for SNe Ia.
This work is made possible by the Supernova Legacy Survey (SNLS)
First Year Data Release \citep{Astier06}.

We present our method in Sec.2, and the results in Sec.3.
Sec.4 contains a simple guide to the use of our photometric
redshift estimator. We discuss and summarize in Sec.5.

\section{The Method}

We derive the empirical photometric redshift estimator
for SNe Ia by using observables that reflect the properties
of SNe Ia as calibrated standard candles.

If SNe Ia were perfect standard candles,
the most important observable in estimating their
redshifts is the peak brightness. Since the SNLS has the 
best sampled light-curves in the $i$ band, we use the $i$
band maximum flux. We use the fluxes in $grz$ at the epoch 
of $i$ maximum flux to make an effective K-correction
to the $i$ flux. Our first estimate of redshift is given by
\be
z_{phot}^{0} = c_1 + c_2 g_f +c_3 r_f + c_4 i_f + c_5 z_f +c_6 i_f^2
\label{eq:z0}
\ee
where $g_f=2.5\log(f_g)$, $r_f=2.5\log(f_r)$, $i_f=2.5\log(f_i)$,
and $z_f=2.5\log(f_z)$, with $f_g$, $f_r$, $f_i$, $f_z$ are fluxes
in ADU counts in $griz$ at the epoch of $i$ maximum flux.

Next, we calibrate each SN Ia in its estimated restframe. 
We define
\be
\Delta i_{15}= 2.5 \log(f_i^{15d}/f_i),
\label{eq:del_i15}
\ee
where $f_i^{15d}$ is the $i$ band flux at 15 days after
the $i$ flux maximum in the estimated restframe, corresponding
to the epoch of $\Delta t^{15d}=15*(1+z_{phot}^{0})$ days after the
epoch of $i$ flux maximum.	    

We now arrive at the final photometric redshift estimator
\be
z_{phot} = c_1 + c_2 g_f +c_3 r_f + c_4 i_f + c_5 z_f +c_6 i_f^2   
		+c_7 \Delta i_{15}
\label{eq:z_a}
\ee

The coefficients $c_i$ (i=1,2,...,7) are found by using
a training set of SNe Ia with $griz$ light-curves and
measured spectroscopic redshifts. We use the jackknife 
technique \citep{Lupton93} to estimate the bias-corrected
mean and the covariance matrix of $c_i$.

\section{Results}

The Supernova Legacy Survey (SNLS) First Year Data Release 
\citep{Astier06} consists of the photometry and redshifts 
of 71 SNe Ia. Of these, only 40 have $griz$ light-curves
with $gz$ photometry covering the epoch of the maximum
flux in the $i$ band. For each of these SN Ia, we fit the
fluxes in the $i$ band light-curve to an asymmetrically stretched
Gaussian introduced by \cite{Wang99}:
\be
f=f_0\,\exp\left\{-\left[ \frac{t-t_{peak}}{w (t-t_{start})^q}
\right]^2 \right\}.
\label{eq:LC fit}
\ee
All the SNe Ia we used are well fitted by this form in the
regions of interest (not too close to the tails).
This yields a smooth light curve without spurious features.
The $i$ band maximum flux and its corresponding epoch are
given by $f_0$ ($f_i=f_0$) and $t_{peak}$ respectively.
The fluxes in $grz$ at the same epoch ($t_{peak}$), $f_g$, $f_r$, $f_z$ 
are obtained from the $grz$ light-curves using linear interpolation. 
We assume a floor of $f_g=200$ (about the size of
the flux errors) for SNe Ia with $f_g <0$.
Eq.(\ref{eq:z0}) is then used to obtain a first estimate
of the SN redshifts, which allows an estimate of $\Delta t^{15d}$.
Eq.(\ref{eq:LC fit}) then gives $f_i^{15d}$ needed for
Eq.(\ref{eq:del_i15}). Eq.(\ref{eq:z_a}) gives the 
final result for the estimated redshifts of the SNe Ia.

Fig.1 shows the photometric redshifts estimated using
the estimator derived in this paper, compared to the
measured spectroscopic redshifts.
The upper panel shows the results for using all 40 SNe Ia
in deriving the coefficients in Eq.(\ref{eq:z_a}).
The rms dispersion in $(z_{phot}-z_{spec})/(1+z_{spec})$
is 0.036. Note that there are no outlyers.
This demonstrates the tight correlation between
the $griz$ fluxes and $\Delta i_{15}$ with redshifts.
The lower panel shows the results for using only the set 
containing the most recently discovered 20 SNe Ia in
deriving the coefficients in Eq.(\ref{eq:z_a}).
These coefficients are then used to predict the redshifts
of the other 20 SNe Ia.
The rms dispersion in $(z_{phot}-z_{spec})/(1+z_{spec})$
is 0.031 for the 20 SNe Ia used in the training set, 
and 0.050 for 20 SNe Ia not used in the training set.

To avoid biases that arise from hand picking the SNe Ia
used in the training set, we have kept the 40 SNe Ia from SNLS
in the order of their discovery, then split them evenly in
the middle into two subsets, and used these as the training
set and testing set respectively for our photometric 
redshift estimator.

Note that the lowest redshift SN Ia in the SNLS $griz$ 
sample is at $z=0.263$. Thus this photometric
redshift estimator is not calibrated for SNe Ia
at $z<0.263$. It will be straightforward to modify
and extend this estimator to lower redshifts, as larger 
uniform samples of SNe Ia covering
a greater range of redshifts become available. 

Peculiar or highly extincted SNe Ia can sometimes
be mistaken as high redshift SNe Ia.
Table 1 lists 5 peculiar and 2 highly extincted SNe Ia
(all are nearby).
It demonstrates that the photometric redshift estimator presented
here generally yields accurate or negative estimated 
redshifts for peculiar and highly extincted SNe Ia.
Thus it will not lead to contamination of the high $z$
sample by nearby peculiar and highly extincted SNe Ia.
\begin{table*}[htb]
\caption{Behavior of the photometric redshift estimator
for peculiar and highly extincted SNe Ia.}
\begin{center}
\begin{tabular}{rrrr}
\hline
SN name & $z_{photo}$ & characteristic & reference \\
\hline
 SN2005hk & -0.14 & peculiar & \cite{Phillips06b} \\
 SN2002cx & -0.12 & peculiar & \cite{Li03}\\
 SN1999ac &  -0.1 & peculiar & \cite{Phillips06a}\\
 SN1999by &  0.00  & peculiar & \cite{Garnavich04}\\
 SN1997br &  -0.01 & peculiar & \cite{Li99}\\
 SN2006X  & -0.01 & highly extincted & \cite{Krisciunas06}\\
 SN1997cy & -0.12 &  highly extincted & \cite{Germany00}\\
 \hline		
\end{tabular}
\end{center}
\end{table*}

\section{A Recipe for Using the Photometric Redshift Estimator}

We now give a practical guide to the use of our photometric
redshift estimator, Eq.(\ref{eq:z_a}). The coefficients $c_i$ have been
derived using 20 SNe Ia, and tested using another 20 SNe Ia
[see Fig.1(b)]. The bias-corrected mean and standard deviations of 
$c_i$, computed using the jackknife technique, are given by
\ba
    c_1 &= & 6.122 \pm 2.006 \nonumber\\
    c_2 &= & -0.06545 \pm 0.04548 \nonumber\\
    c_3 &= & -0.03268 \pm 0.08201 \nonumber\\
    c_4 &= & -0.8225 \pm 0.4513 \nonumber\\
    c_5 &= & -0.06292 \pm 0.08601 \nonumber\\
    c_6 &= &  0.03979 \pm 0.02104 \nonumber\\
    c_7 &= &   0.04552 \pm 0.04335
    \label{eq:c_i}
\ea
We give the covariance matrix of $c_i$ in Table 2.

\begin{table*}[htb]
\caption{The covariance matrix of $c_i$.}
\begin{center}
\begin{tabular}{rrrrrrr}
\hline\hline
  0.4022E+01 & 0.3364E-02 & 0.6515E-01 &-0.8817E+00 & 0.2031E-01 & 0.3988E-01 & 0.3912E-01\\
  0.3364E-02 & 0.2068E-02 & 0.3714E-03 & 0.2112E-02 &-0.2602E-02 &-0.2421E-03 & 0.9723E-03\\
  0.6515E-01 & 0.3714E-03 & 0.6726E-02 &-0.1750E-01 & 0.1978E-03 & 0.4210E-03 & 0.1924E-03\\
 -0.8817E+00 & 0.2112E-02 &-0.1750E-01 & 0.2037E+00 &-0.1203E-01 &-0.9101E-02 &-0.5894E-02\\
  0.2031E-01 &-0.2602E-02 & 0.1978E-03 &-0.1203E-01 & 0.7398E-02 & 0.5995E-03 &-0.1549E-02\\
  0.3988E-01 &-0.2421E-03 & 0.4210E-03 &-0.9101E-02 & 0.5995E-03 & 0.4429E-03 & 0.2314E-03\\
  0.3912E-01 & 0.9723E-03 & 0.1924E-03 &-0.5894E-02 &-0.1549E-02 & 0.2314E-03 & 0.1879E-02\\
\hline
 \hline		
\end{tabular}
\end{center}
\end{table*}

Here are the steps one should follow in using our photometric redshift estimator:
(1) Estimate the flux and epoch of the $i$ band maximum, $f_i$ and $t_{peak}$.
(2) Estimate the fluxes in $grz$ at the same epoch, $f_g$, $f_r$, and $f_z$.
(3) Use Eq.(\ref{eq:z_a}) with $c_i$ (i=1,2,...,6) given in Eq.(\ref{eq:c_i})
and $c_7=0$ to obtain a first estimate of $z_{phot}^0$.
(4) Use $z_{phot}^0$ to estimate $\Delta t^{15d}=15*(1+z_{phot}^{0})$.
(5) Estimate the $i$ band flux at $\Delta t^{15d}$ days after $t_{peak}$, $f_i^{15d}$,
and compute $\Delta i_{15}= 2.5 \log(f_i^{15d}/f_i)$.
(6) Use Eq.(\ref{eq:z_a}) with $c_i$ (i=1,2,...,7) given in Eq.(\ref{eq:c_i})
to obtain $z_{phot}$.
(7) Use the covariance matrix of $c_i$ given in Table 1 to 
compute the standard deviation of $z_{phot}$.

Note that the SNLS zeropoints should be used to convert 
$g_f$, $r_f$, $i_f$, $z_f$ in Eqs.(\ref{eq:z0}) and (\ref{eq:z_a}) 
into magnitudes \citep{Astier06}, and the appropriate conversions
should be made if the available photometry is in $BVRI$ instead 
of $griz$ \citep{Fuku96}.

\section{Discussion and Summary}

We have derived a model-independent photometric redshift estimator, 
Eqs.(\ref{eq:z_a}) \& (\ref{eq:c_i}), for SNe Ia that uses only 
multiband photometry near maximum light, 
and a training set of SNe Ia
with multiband photometry and measured spectroscopic redshifts.
This estimator is simple and analytical, thus is very easy to
implement (see Sec.4).  

The test of our photometric redshift estimator using SNe Ia not 
used in the training set demonstrates that this estimator is robust, 
and with an accuracy that is of the same order of
magnitude as the photometric errors (see Fig.1(b)).
The large uncertainties in the coefficients $c_i$ used in 
our photometric redshift estimator (see Eq.(\ref{eq:c_i})
are indicative of the relatively small sample (20 SNe Ia) used for
the training set, as well as photometric errors.
Eq.(\ref{eq:c_i}) also shows that the most key constraints
on redshift comes from the $g$ and $i$ band photometry.

As the quality of data improves, we expect that our photometric
redshift estimator can be further improved to provide more
accurate redshift estimates. 
This photometric redshift estimator can be easily modified
and trained to apply to SNe Ia photometry in any choices
of multiple bands.

The measurement of spectroscopic redshifts of SNe is the most
costly and constraining aspect of a supernova survey.
Our results will allow observers to estimate the redshifts
rather accurately based on near maximum 
light multiband photometry only (after obtaining spectroscopic redshifts
for a modest training set), thus greatly increase the efficiency
of supernova spectroscopy of high redshift candidates.

In order to model the systematic uncertainties of SNe Ia
as standard candles, it is critical to obtain a very large
number of SNe. Future supernova surveys can easily obtain 
the multiband photometry of a huge number of supernovae
\citep{Wang00,JEDI}, for example, using the
Advanced Liquid-mirror Probe for Astrophysics, Cosmology and 
Asteroids (ALPACA) \footnote{http://www.astro.ubc.ca/LMT/alpaca/}, 
and the Large Synoptic Survey Telescope (LSST) \footnote{http://www.lsst.org/}.
It will not be practical to obtain spectroscopic redshifts
for all the SNe Ia found by such surveys. 
With the dense sampling and accurate multiband photometry expected 
for future supernova surveys, it will be possible to refine
our photometric redshift estimator to the accuracies suitable
for cosmology \citep{Huterer04}.
This will dramatically boost the cosmological impact 
of very large future supernova surveys.

\bigskip

{\bf Acknowledgements}
I am grateful to the SNLS team for releasing their data,
including multi-band light-curves in flux units, and
to Kevin Krisciunas for providing me with unpublished
photometry on SN2006X.
This work is supported in part by NSF CAREER grant AST-0094335.

\clearpage
\setcounter{figure}{0}

\begin{figure}
\centering
\includegraphics[width=14cm]{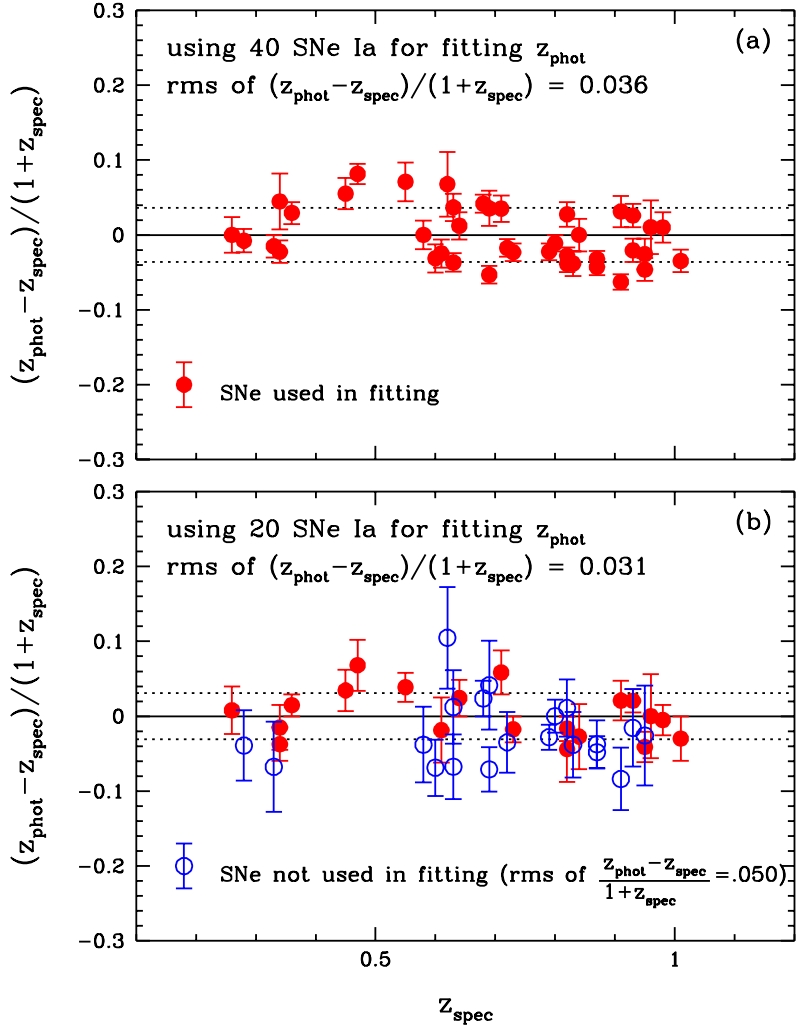}
\caption{The photometric redshifts estimated using
the estimator in Eq.(\ref{eq:z_a}), compared to the
measured spectroscopic redshifts.
(a) The results for using all 40 SNe Ia
in deriving the coefficients in Eq.(\ref{eq:z_a}).
(b) The results for using only the set 
containing the most recently discovered 20 SNe Ia in
deriving the coefficients in Eq.(\ref{eq:z_a}).
These coefficients are then used to predict the redshifts
of the other 20 SNe Ia.
}
\label{fig1}
\end{figure}

\end{document}